\documentclass[12pt]{iopart}

\usepackage{booktabs}
\usepackage[colorlinks = true,
            linkcolor = cyan,
            urlcolor  = blue,
            citecolor = blue,
            anchorcolor = black]{hyperref}	
\usepackage{soul} 
\usepackage{setspace}
\usepackage{amssymb}
\usepackage{graphicx}
\usepackage[labelsep=period]{caption}
\usepackage{float}

\usepackage[dvipsnames]{xcolor}

\usepackage{soul}

\makeatletter
\newcommand{\smallsym}[2]{#1{\mathpalette\make@small@sym{#2}}}
\newcommand{\make@small@sym}[2]{%
  \vcenter{\hbox{$\m@th\downgrade@style#1#2$}}%
}
\newcommand{\downgrade@style}[1]{%
  \ifx#1\displaystyle\scriptstyle\else
    \ifx#1\textstyle\scriptstyle\else
      \scriptscriptstyle
  \fi\fi
}
\makeatother

\begin{document}

\title{Invariant regimes of Spencer scaling law for magnetic compression of rotating FRC plasma}

\author{Yiming Ma$^1$, Ping Zhu$^{1,2*}$, Bo Rao$^{1*}$ and Haolong Li$^3$}

\address{
State Key Laboratory of Advanced Electromagnetic Technology, International Joint Research Laboratory of Magnetic Confinement Fusion and Plasma Physics, School of Electrical and Electronic Engineering, Huazhong University of Science and Technology, Wuhan, 430074, China
\\
$2$ Department of Nuclear Engineering and Engineering Physics, University of Wisconsin-Madison, Madison, Wisconsin 53706, USA \\
$3$ College of Physics and Optoelectronic Engineering, Shenzhen University, Shenzhen 518060, China

}
\ead{zhup@hust.edu.cn, borao@hust.edu.cn}

\vspace{10pt}


\begin{abstract}

The scaling laws for the magnetic compression of a toroidally rotating field reversed configuration (FRC) have been investigated in this work. The magnetohydrodynamics (MHD) simulations of the magnetic compression on rotating FRCs employing the NIMROD code [C. R. Sovinec \textit{et al.}, J. Comput. Phys. \textbf{195}, 355 (2004)], are compared with the Spencer's one-dimensional (1D) theory [R. L. Spencer \textit{et al.}, Phys. Fluids \textbf{26}, 1564 (1983)] for a wide range of initial flow speeds and profiles. The toroidal flow can influence the scalings directly through the alteration of the compressional work as also evidenced in the 1D adiabatic model, and indirectly by reshaping the initial equilibrium. However, in comparison to the static initial FRC equilibrium cases, the pressure and the radius scalings remain invariant for the magnetic compression ratio $B_{w2}/B_{w1}$ up to 6 in presence of the initial equilibrium flow, suggesting a broader applicable regime of the Spencer scaling law for FRC magnetic compression. The invariant scaling has been proven a natural consequence of the conservation of angular momentum of both fluid and magnetic field during the dynamic compression process.

\end{abstract}

\vspace{0.5pc}

\section{Introduction}

The field reversed configuration (FRC) is a high-$\beta$ compact toroid primarily sustained by the poloidal magnetic field generated from plasma diamagnetic currents~\cite{tuszewski1988field, rej1986experimental}. As a promising candidate path to compact fusion, the FRC has the linear device geometry, the extended lifetime surpassing magnetohydrodynamics (MHD) expectations~\cite{schwarzmeier1983magnetohydrodynamic, slough1995transport, steinhauer1996frc}, and the capability to withstand highly dynamic processes such as supersonic translation and collision merging~\cite{liao2022first, liao2022dynamic, sekiguchi2018super, kobayashi2021experimental, asai2021observation, asai2019collisional, gota2021overview}. The magnetic compression over a short timescale (tens of microseconds) has been demonstrated as an effective heating method for FRCs in experiments~\cite{slough2011creation, steinhauer2011review, rej1992high, kirtley2023fundamental}. The scaling predictions from the one-dimensional (1D) adiabatic compression model by Spencer~\cite{spencer1983adiabatic} has been compared with FRX-C/LSM (Field-Reversed Experiment-C/Large `S'
Modification) experiments~\cite{rej1992high} and MHD simulations~\cite{woodruff2008adiabatic, ma2023mhd}. Starting from static FRC equilibria in absence of any toroidal flow, previous simulation results~\cite{ma2023mhd} agree with the experiments~\cite{rej1992high} qualitatively while confirming some key scaling and geometry relations in the analytical model~\cite{spencer1983adiabatic}, despite the fact that in reality, FRCs often have strong toroidal rotation at the order of $0.1V_A$~\cite{steinhauer2011review, belova2005numerical, belova2001numerical, belova2003kinetic, belova2004kinetic, harris2009ion, ono2003spontaneous}. 
The degree to which this agreement might be coincidental, the mechanisms by which toroidal flow may influence the scalings, and the potential existence of invariant scaling regimes in presence of flow all remain unclear.

In this work, we investigate these issues in a set of new simulations of the magnetic compression processes starting from initially rotating FRC equilibria using the resistive MHD model implemented in the NIMROD code~\cite{glasser1999nimrod, sovinec2003nimrod, sovinec2004nonlinear}. The toroidal flow effects  have been studied using a variety of steady equilibria characterized by various flow profiles and speeds as well as density profiles. The invariant regimes have been identified where the scalings for magnetic compression remain almost unchanged even for the initially rotating FRCs. It turns out that such an invariance of scaling originates from the conservation of angular momentum during the dynamic compression process.

The rest of paper is organized as follows. After a brief discussion on the 1D adiabatic model in section~\ref{sec: Adiabatic compression scaling}, the numerical approach and simulation results are described in section~\ref{sec: Numerical simulation}. Finally, the discussion and conclusions are presented in section~\ref{sec: Discussion and conclusions}.

\section{Analytical model}

\subsection{1D adiabatic compression model}
\label{sec: Adiabatic compression scaling}

A schematic plot of the FRC is shown as figure~\ref{fig: frc scheme and eq rr}(a). For the highly elongated FRCs in absence of initial equilibrium flow, the quasi-static adiabatic compression yields $dW=dE$, where the $dW$ is the external work done on the separatrix and $dE$ is the increment of the total FRC energy $E = \int \left ( p / (\gamma -1) + B^2 / (2\mu_0) \right) dV$. Considering the static equilibrium force balance
\begin{equation}
  \label{eqn: 1d eq}
  p_m = p(\psi) + \frac{B_z^2(\psi)}{2\mu_0} = \frac{B_w^2}{2\mu_0}
\end{equation}
along with the field line structural relation $r_s = \sqrt{2} r_o$ and the axial force balance condition $\langle \beta \rangle = 1 - x_s^2 /2 $~\cite{armstrong1981field, lee2020generalized}, the adiabatic compression scaling laws for both the magnetic and wall compressions can be obtained from the adiabatic relation $dW=dE$~\cite{spencer1983adiabatic, intrator2008adiabatic}, where $p_m$ is the maximum plasma pressure, $B_w$ is the magnetic field outside the FRC plasma, $r_s$ is the major radius of separatrix, $r_o$ is the radius of the magnetic axis or O-point, $\langle \beta \rangle$ is the average plasma $\beta$ over the middle plane within the separatrix, $x_s = r_s / r_w$ is the ratio of separatrix radius to the wall radius, $\psi$ is the poloidal flux and the cylindrical coordinates $(r, \theta, z)$ is adopted. The Spencer scaling laws for magnetic compression~\cite{spencer1983adiabatic} are listed in table~\ref{tab: FRC scaling}, where $l$ is the length of the FRC, $\epsilon$ is a flux profile index, $n_m$ is the maximum number density and $T_m$ represents the maximum total plasma temperature. The $\epsilon=0.25$ is adopted in this study in line with previous scaling law applications~\cite{intrator2008physics, tuszewski1988semiempirical}.

\begin{table}[H]
  \caption{Spencer's scaling laws~\cite{spencer1983adiabatic} for magnetic compression, $\epsilon > 0$ and $\gamma=5/3$}
  \label{tab: FRC scaling}
  \centering
   \begin{tabular}{p{4cm}p{4cm}}
     \toprule
     parameter  & scaling law \\
     \cmidrule(r){1-2}

     $B_w$ & $ x_s^{-(3+\epsilon)}$\\

     $p_m$ & $ x_s^{-2(3+\epsilon)}$ \\

     $l$  & $  x_s^{ \frac{ 2(4+3\epsilon) }{ 5 } }  \langle \beta \rangle^{ \frac{-(3+2\epsilon)}{5} } $ \\

     $n_m$  & $ x_s^{\frac{ -6(3+\epsilon) }{ 5 }} \langle \beta \rangle^{\frac{ -2(1-\epsilon) }{ 5 }} $  \\

     $T_m$  & $ x_s^{\frac{-4(3+\epsilon)}{5}} \langle \beta \rangle^{ \frac{ 2(1-\epsilon) }{ 5 } } $ \\

     \bottomrule
   \end{tabular}
\end{table}

The flow effects can be preliminarily assessed within the adiabatic model and subsequently subjected to quantitative analysis through 2D MHD simulations. The force balance in the 1D  steady FRC equilibrium is described by the equation
\begin{equation}
  \label{eqn: 1D eq}
  \frac{d}{dr} \left(p + \frac{B_z^2}{2\mu_0} \right) = \rho \Omega^2 r
\end{equation}
where $\rho$ is the mass density and $\Omega$ is the toroidal or azimuthal rotation frequency. Upon integrating equation~(\ref{eqn: 1D eq}) and assuming 
the rotation and density profiles from the rigid rotor (RR) model, equation~(\ref{eqn: 1D eq}) can be simplified as
\begin{equation}
  \label{eqn: 1D eq 2}
  p + \frac{B^2}{2\mu_0} = \frac{m_i n_m \Omega^2 r_s^2  \tanh(k u)}{4k}  + p_m 
\end{equation}
with the relation $r_s = \sqrt{2} r_o$ and the condition $p(r=r_o) = p_m$. More specifically the rotation frequency $\Omega = u_\theta /r = \mathrm{const}$ and
\begin{equation}
  \label{eqn: rr density profile}
  n=n_m \mathrm{sech}^2 (k u)
\end{equation}
are adopted in the RR model, where $k$ is the shape factor and the minor radius variable $u=2r^2/r_s^2-1$. The total FRC energy $E = \int \left( p/(\gamma-1) + B^2/(2\mu_0) + \rho \Omega^2 r^2 /2 \right) dV$ then can be written as
\begin{equation}
  \label{eqn: total en with flow}
  E = p_m V(1+ \frac{1}{2} \langle \beta \rangle ) + \frac{1}{4} m_i n_m \Omega^2 r_s^2 \langle \beta \rangle V
\end{equation}
after taking into account equation~(\ref{eqn: 1D eq 2}) and $\langle \beta \rangle = \tanh(k) / k$.
The first term of equation~(\ref{eqn: total en with flow}) comprises the internal energy and the energy stored in magnetic field, while the second term accounts for the kinetic energy. A rise of the rotation frequency is anticipated during the compression, and the $\Omega \propto x_s^{-2}$ relation can be deduced by assuming the plasma angular momentum alone remains unchanged in conjunction with the RR flow and density profile. In particular, the fluid inventory conservation $dN = 0$ provides
\begin{equation}
  \label{eqn: dN=0}
  2\frac{dr_s}{r_s} + \frac{dl}{l} + \frac{dn_m}{n_m} + \frac{d\langle \beta \rangle}{\langle \beta \rangle} = 0 \, ,
\end{equation}
and  $dL = 0$ yields
\begin{equation}
  \label{eqn: dL=0}
  \frac{d\Omega}{\Omega} + 4\frac{dr_s}{r_s} + \frac{dl}{l} + \frac{dn_m}{n_m} + \frac{d\langle \beta \rangle}{\langle \beta \rangle} = 0
\end{equation}
where the angular momentum $L=\int \rho \Omega r^2 dV =m_i n_m \Omega r_s^2 \langle \beta \rangle V /2$, $N = \int n dV =  n_m \langle \beta \rangle V$ and $V=\pi r_s^2 l$. Consequently $d\Omega / \Omega + 2 dr_s / r_s = 0$ can be obtained from combining equation~(\ref{eqn: dN=0}) and equation~(\ref{eqn: dL=0}).  Due to the conversion of a fraction of work done on the separatrix into kinetic energy during the compression, the pressure is expected to reduce relatively to the situation in absence of initial flow, thereby leading to a less effective compressional heating. It is noted that while $dL=0$ holds as the FRC maintains steady equilibrium during compression, this condition is not met in the dynamic compression process when the FRC deviates from its equilibrium state, as discussed next in section~\ref{sec: L conservation}.

\subsection{Angular momentum conservation in dynamic compression}
\label{sec: L conservation}

In a toroidally rotating plasma system in absence of any direct deposition from external angular momentum source such as the neutral beam injection, the angular momentum conservation is governed by the equation:
\begin{equation}
    \label{eqn: 01}
    \frac{dL}{dt} = \tau_{z, tot} 
\end{equation}
and 
\begin{equation}
    \tau_{z, tot} = -\int  r\hat{\theta} \cdot (\nabla \cdot \stackrel{\leftrightarrow}{T}) dV ,
\end{equation}
where $L = \int r \rho u_\theta dV$ is the axial angular momentum of plasma and $\tau_{z, tot}$ represents the total axial torque~\cite{pustovitov2011integral}. $\stackrel{\leftrightarrow}{T}$ is the stress tensor
\begin{equation}
    \stackrel{\leftrightarrow}{T} = \rho \vec{u}\vec{u} + (p + \frac{B^2}{2\mu_0}) \stackrel{\leftrightarrow}{I} - \frac{\vec{B} \vec{B}}{\mu_0} + \stackrel{\leftrightarrow}{\Pi} ,
\end{equation}
$\stackrel{\leftrightarrow}{\Pi}$ is the viscosity tensor and $\stackrel{\leftrightarrow}{I}$ is the unit tensor. During the dynamic compression, the total torque is primarily composed of the electromagnetic torque $\tau_{z,EM}$ corresponding to the Maxwell stress tensor $\stackrel{\leftrightarrow}{T\!}_M = -B^2/(2\mu_0) \stackrel{\leftrightarrow}{I} + \vec{B} \vec{B}/\mu_0$ and the torque $\tau_{z,R}$ induced by the Reynold stress tensor $\stackrel{\leftrightarrow}{T\!}_R = \rho \vec{u} \vec{u}$
\begin{equation}
\label{eqn: tau tot approx}
    \tau_{z,tot} \approx \tau_{z,EM} + \tau_{z,R} ,
\end{equation}
where
\begin{equation}
\label{eqn: tau EM}
    \tau_{z,EM} = \int \frac{r}{\mu_0} B_\theta \vec{B}\cdot \vec{n} dS
\end{equation}
and
\begin{equation}
    \tau_{z,R} = -\int r\rho u_\theta \vec{u} \cdot \vec{n} dS 
\end{equation}
within a cylindrical volume enclosed by the surface with unit normal vector $\vec{n}$. Due to the dynamic nature of the compression process, the nonzero $B_\theta$ driven by the differential rotation with nonuniform rotation frequency $\Omega$~\cite{guzdar1985role}, and the finite radial compression velocity $\vec{u}\cdot\vec{n}$ can both enable violation of the condition $dL=0$ that is assumed in the 1D quasi-static adiabatic compression model in the previous subsection~\ref{sec: Adiabatic compression scaling}. The consequence of the dynamic angular momentum conservation with $dL \neq 0$ will be demonstrated and analyzed along with the simulation results presented in section~\ref{sec: Numerical simulation}.

\section{Numerical simulation}
\label{sec: Numerical simulation}

\subsection{Simulation model and setup}

The simulations in this work are conducted using the NIMROD code~\cite{sovinec2004nonlinear}, which employs the finite element discretization in the $(r,z)$ plane and Fourier series decomposition in the $\theta$ direction.  To compare with Spencer's scaling laws, the resistive single-fluid MHD equations are solved

\begin{equation}
  \frac{\partial n}{\partial t} + \nabla \cdot (n\vec{u})= 0 \, ,
\end{equation}

\begin{equation}
  \rho(\frac{\partial \vec{u}}{\partial t} + \vec{u}\cdot \nabla\vec{u} ) = \vec{J}\times \vec{B} - \nabla p - \nabla \cdot \stackrel{\leftrightarrow}{\Pi} \, ,
\end{equation}

\begin{equation}
  \label{eqn: mhd temperature}
\frac{n}{\gamma-1} (\frac{\partial}{\partial t} + \vec{u}\cdot \nabla ) T = -p \nabla \cdot \vec{u} + Q \, ,
\end{equation}

\begin{equation}
  \frac{\partial \vec{B}}{\partial t} = - \nabla \times \vec{E} \, ,
\end{equation}

\begin{equation}
  \vec{E} = -\vec{u}\times \vec{B} + \eta \vec{J} \, ,
\end{equation}
where $n$ is the plasma number density, $\vec{u}$ the plasma velocity, $\rho$ the mass density, $\vec{J}$ the current density, $\vec{B}$ the magnetic field, $p$ the pressure, $\stackrel{\leftrightarrow}{\Pi}$ the viscosity tensor, $T=T_i+T_e$ the total temperature, $\vec{E}$ the electric field, $\eta$ the resistivity and $Q=\eta J^2$ represents the resistive heating power.  Employing the Spitzer's resistivity model, a core plasma resistivity corresponding to a Lundquist number of $6\times 10^4$ is initialized, with the wall radius adopted as the relevant length scale. Minimal values of viscosity and density diffusivity are kept to ensure the numerical stability and convergence of the simulation results.   An isotropic viscosity corresponding to the Reynolds number of $2\times 10^4$ is applied. A parallel viscosity of $10^4 \mathrm{m^2 / s}$ is used as well following previous FRC simulations~\cite{milroy2010extended, macnab2007hall,ma2023mhd}. Moreover, a supplementary ohmic heating introduced in the simulations is two orders of magnitude lower than the internal energy.  The magnetic compression boundary condition is externally applied on the $r=r_w$ boundary, where the compression field $B_z$ ramps with time~\cite{ma2023mhd}.

The initial FRC equilibrium before the compression is prepared by numerically solving the Grad-Shafranov (GS) equation with toroidal rotation, using the extended NIMEQ code~\cite{li2021solving,howell2014solving}. The equilibrium exhibits variations in pressure and magnetic flux distributions due to the toroidal flow effects~\cite{evangelias2016axisymmetric, guazzotto2009magnetic}, where the pressure and the mass density are functions of both $\psi$ and $r$~\cite{li2021solving, li2021formation, aiba2009minerva}
\begin{equation}
  \label{eqn: pres wi flow}
  p(\psi, r)=p_0(\psi) \exp \left[\frac{m_i r_o^2 \Omega^2}{2 T}\left(\frac{r^2}{r_o^2}-1\right)\right],
\end{equation}

\begin{equation}
  \rho(\psi, r)=\rho_0(\psi) \exp \left[\frac{m_i r_o^2 \Omega^2}{2 T}\left(\frac{r^2}{r_o^2}-1\right)\right],
\end{equation}
with $p_0(\psi)$ and $\rho_0(\psi)$ being the static equilibrium pressure and mass density profiles respectively. The initial equilibrium flow is thus anticipated to have an indirect impact on the compression process through the modifications to initial equilibrium itself. Detailed analyses of flow effects on the compression will be conducted in subsequent simulations in the range of $\alpha < 1.5 - 2$, where  $\alpha=\langle u_{\theta} \rangle / u_{*}$~\cite{harned1984origin, belova2000numerical} and $u_{*} = \langle J / (e n) \rangle$ are averages over the middle plane. This range of equilibrium flow speeds encompass the regime observed in the majority of FRC experiments and simulations~(e.g.~\cite{belova2005numerical, belova2004kinetic, ono2003spontaneous, ito1987ion}). The influence of different toroidal rotation profiles on the compression scaling laws is also studied, including the rigid rotation, a single-peaked rotation profile with the maximum flow speed near the O-point~\cite{ono2003spontaneous}, and a double-peaked rotation profile similar to that in Ref.~\cite{belova2000numerical}, using a cubic function and a Gaussian shaped function for $\Omega(\psi)$ respectively~\cite{howell2014solving, li2021formation}.

\subsection{Effects of initial rigid rotation}
\label{subsec: RR}

\begin{figure*}[!htbp]
  \centering
  \includegraphics[width=0.8\textwidth]{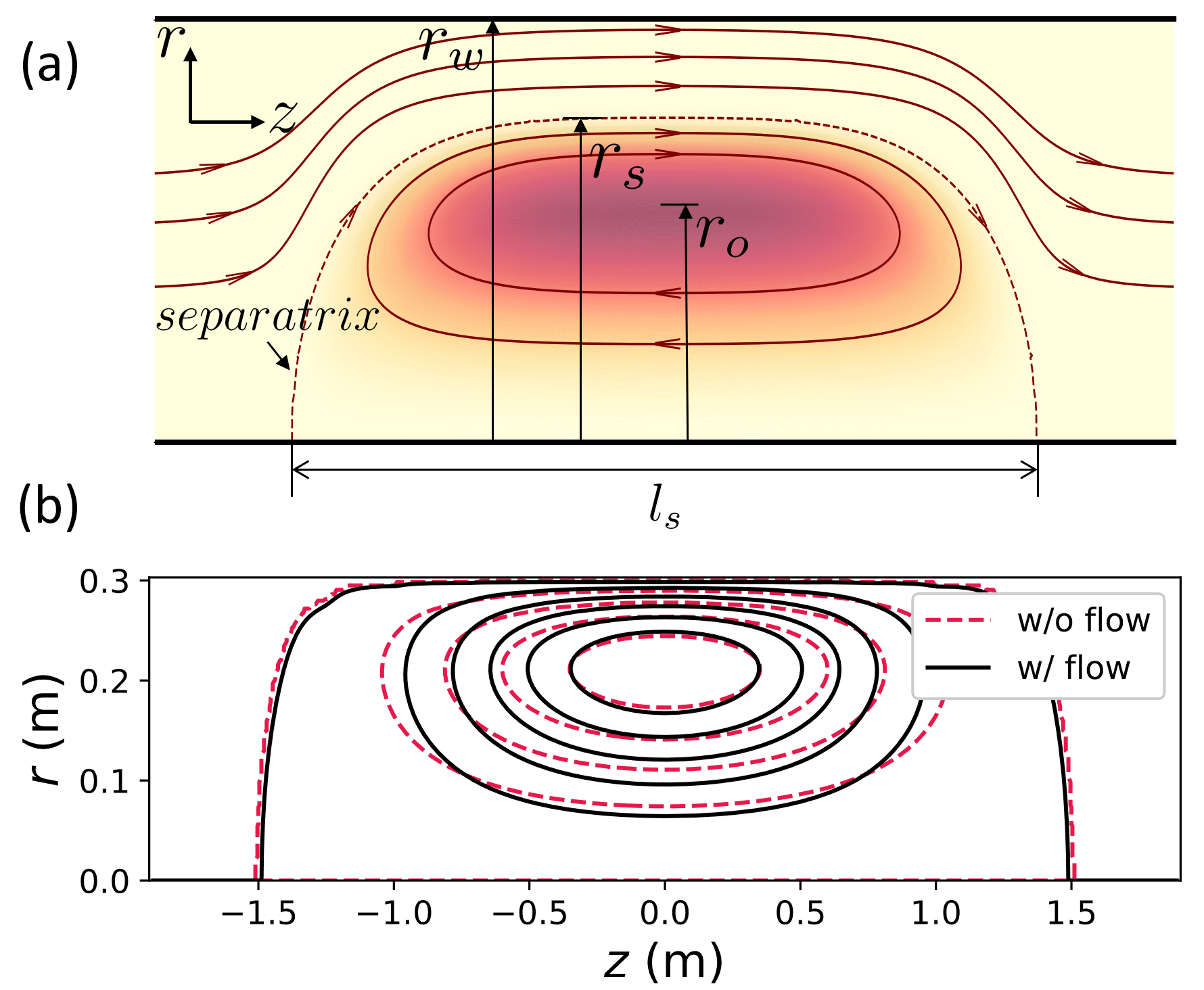}
  \caption{(a) A schematic plot of the FRC, where $r_o$, $r_s$, and $r_w$ denote the radii of the magnetic axis, the separatrix, and the wall locations respectively, and $l$ denotes the length of the FRC defined as the distance between the two separatrix-axis intersections. (b) The poloidal flux contours in the equilibriums in absence of flow (red) and in presence of the RR flow (black) respectively.}
  \label{fig: frc scheme and eq rr}
\end{figure*}

We first compare the simulation results on the magnetic compression scalings between a static initial equilibrium and those with an initial rigid rotation.  The simulations in absence of an initial flow are performed following the earlier simulation setups~\cite{ma2023mhd}, and the corresponding equilibrium magnetic flux contours are presented in figure~\ref{fig: frc scheme and eq rr}(b).  The initial steady equilibrium shares the same $p_0(\psi)$ shown in equation~(\ref{eqn: pres wi flow}) as the static equilibrium when solving the GS equation, where $p_0(\psi)$ is constant outside the separatrix. In magnetic compression experiments a large initial $x_s$ is preferred for high efficiency~\cite{slough2011creation}, and $x_s \sim 1$ is adopted as an optimal case in the simulations. The maximum equilibrium rotation frequency $\Omega_{0m} \sim 0.23\Omega_A$ adopted in simulations corresponds to $\alpha\sim 1.5$ under the RR flow, where $\Omega_A=V_A/r$, $\Omega_{0m}/\Omega_A=V_{0m}/V_A$, $V_{0m}$ is the initial maximum magnitude of the toroidal veolocity $V_t$, which is in the co-rotating direction counter to the $\theta$-direction, i.e. $V_t = -u_\theta$, and $V_A=B_w / (\mu_0 m_i n_m)^{1/2}$ is the characteristic Alfv\'{e}n velocity of the initial equilibrium. The RR flow slightly alters the initial equilibrium flux shape with this flow speed~(figure~\ref{fig: frc scheme and eq rr}(b)). Moreover, the standard calculations are only limited to 2D because the 2D simulations yield the same results as the 3D cases when no initial perturbations are applied to the internal simulation domain, thereby confirming the numerical accuracy of the simulations.

\begin{figure*}[!htbp]
  \centering
  \includegraphics[width=1.0\textwidth]{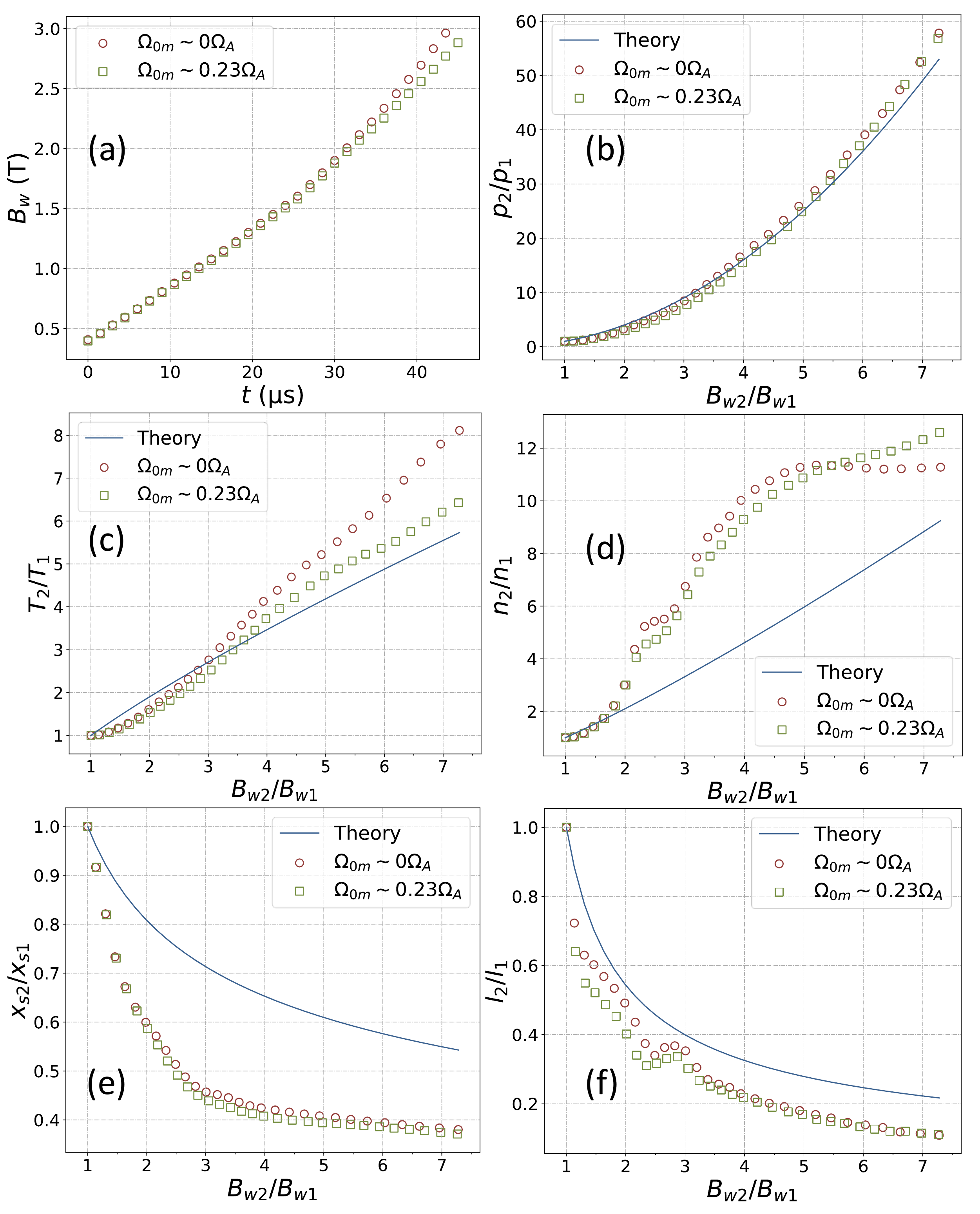}
  \caption{ (a) The compression field $B_w$ variation with time, the (b) pressure ratio $p_2/p_1$, (c) temperature ratio $T_2/T_1$, (d) density ratio $n_2/n_1$, (e) radius ratio $x_{s2}/x_{s1}$ and (f) length ratio $l_2/l_1$ as functions of the magnetic compression ratio $B_{w2}/B_{w1}$,  obtained from Spencer's theory (blue lines) and the two simulation cases: one without flow (red circles) and the other with the initial rigid rotation (green rectangles). }
  \label{fig: RR results}
\end{figure*}

\begin{figure*}[!htbp]
  \centering
  \includegraphics[width=0.9\textwidth]{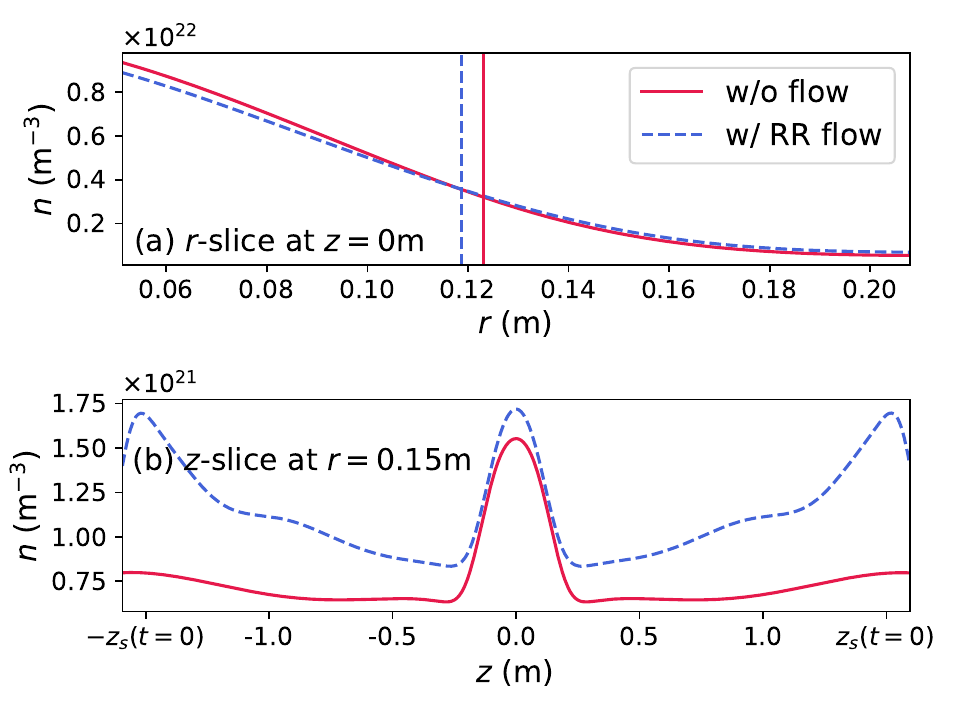}
  \caption{The density profiles at $t=30 \mathrm{\mu s}$ along (a) the $z=0$ middle plane and (b) the $r=0.15\mathrm{m}$ axis. Vertical lines indicate $r=r_s$ for cases in absence of initial flow and with the initial rigid rotation. The simulations are same as the cases shown in figure~\ref{fig: RR results}.}
  \label{fig: centrif rr slice}
\end{figure*}

The $p_m$, $T_m$ and $n_m$ at the $z=0$ middle plane during the compression process in simulations are plotted in figure~\ref{fig: RR results} for comparison with the 1D scalings shown in table~\ref{tab: FRC scaling}. The subscripts ``1" and ``2" denote the equilibrium state before compression and the final state at the corresponding compression ratio respectively. As can be seen from figure~\ref{fig: RR results}, the presence of an initial rigid rotation is responsible for a reduction in the maximum temperature of the compressed FRC. Both the pressure and density are also reduced within the regime $B_{w2}/B_{w1}<5$ in comparison to the case with the static equilibrium. Figure~\ref{fig: centrif rr slice} shows that at $t=30\mathrm{\mu s}$ with $B_{w2}/B_{w1}\sim 4.7$, the presence of initial RR flow leads to a higher density outside the separatrix and slightly lower density inside relative to the case in absence of initial toroidal flow. The centrifugal force can retard the radial inward concentration of the density during compression, which is similarly observed in the tilt mode simulations~\cite{milroy1989nonlinear}. The compressed FRC gradually deviates from the elongated separatrix shape assumed in the 1D adiabatic model during compression, and the simulation cases with and without initial flow show negligible difference in the pressure scaling after $B_{w2}/B_{w1} > 5$. The conversion of a fraction of the compressional magnetic energy into the toroidal kinetic energy leads to overall less effective heating. Consequently the FRC with initially rigid rotation (figure~\ref{fig: RR results}) maintains a poloidal flux approximately $7\%$ lower than that without initial flow at $B_{w2}/B_{w1}\sim 8$. Since $B_w$ is the sum of the externally applied field and that induced by plasma, the $B_w$ shown in figure~\ref{fig: RR results}(a) exhibits discrepancy between the two simulation cases despite identical boundary conditions. The axial contraction of the FRC with initial rigid rotation is slightly faster before $B_{w2}/B_{w1} < 3$. For the other two initial radial profiles of toroidal rotation shown in figure~\ref{fig: eq flow}, the axial contraction of the FRC during compression is somewhat slower than the case without initial flow. The two simulation cases in figure~\ref{fig: RR results} have a peaked equilibrium density profile with $n(r=0)/n_m \sim 0.5$, resembling the RR density profile shown in equation~(\ref{eqn: rr density profile}) with $k=0.9$. Whereas different density profiles have been demonstrated to primarily affect the density and temperature scalings for the magnetic compression of the initially static FRC equilibrium~\cite{ma2023mhd}, the toroidal flow effects induced by an initially rigid rotation are found similar on scalings for both the uniform and the peaked initial density profiles.

\begin{figure*}[!htbp]
  \centering
  \includegraphics[width=0.8\textwidth]{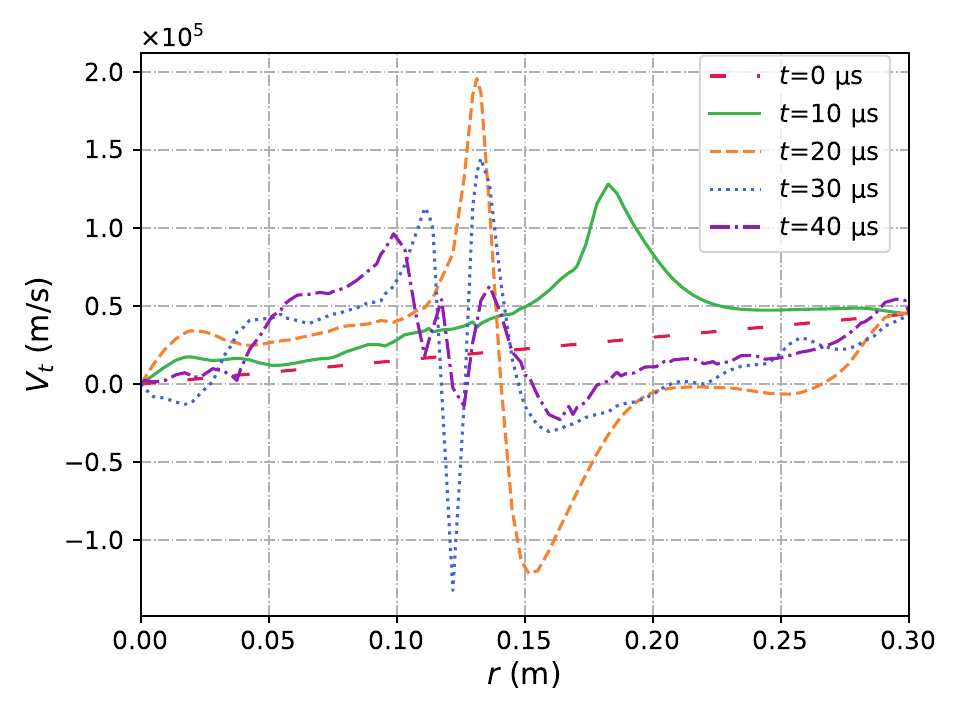}
  \caption{Radial profiles of toroidal velocity along the $z=0$ middle plane at different times during compression. The simulation with initial rigid rotation is same as the case shown in figure~\ref{fig: RR results}.}
  \label{fig: vt slice nu rr}
\end{figure*}

The quasi-static 1D model in section~\ref{sec: Adiabatic compression scaling} predicts a monotonic rise in toroidal rotation and kinetic energy, thus the consequent compression scalings are expected to significantly deviate from the Spencer theory~\cite{spencer1983adiabatic}.  However, in simulations the maximum toroidal velocity, along with the toroidal flow induced deviations in pressure and length scalings from the Spencer theory initially rises but subsequently decreases as shown in figures~\ref{fig: RR results} and~\ref{fig: vt slice nu rr}. This result may be understood by the mitigating effects of the electromagnetic torque in equation~(\ref{eqn: tau EM}) as explained next.

\begin{figure*}[!htbp]
  \centering
  \includegraphics[width=0.8\textwidth]{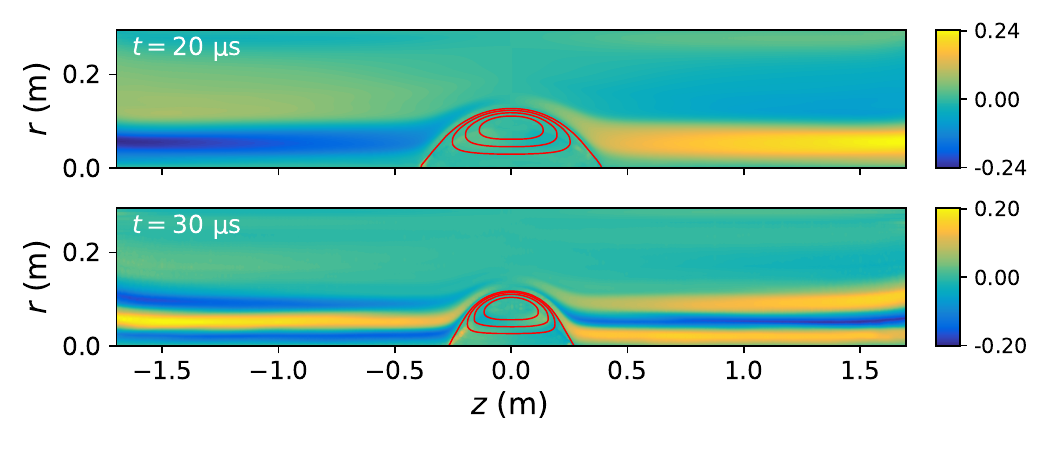}
  \caption{Contours of $B_\theta$ (colored) and poloidal flux (red) at $t=20\mathrm{\mu s}$ and $t=30\mathrm{\mu s}$ for the case with the initial rigid rotation. The simulation is same as the case shown in figure~\ref{fig: RR results}.}
  \label{fig: Btheta contour rr nu 0.23Va}
\end{figure*}

It is noted that a significant $B_\theta$ arises during the compression as figure~\ref{fig: Btheta contour rr nu 0.23Va} demonstrates. For the cylinder with a length between $-z_{end}$ and $z_{end}$ and a radius of $r_{end}$, the axial electromagnetic torque from each end surface of the cylinder can be expressed as
\begin{equation}
    \tau_{1} = \int_0^{r_{end}} \frac{2\pi}{\mu_0} r^2 B_\theta B_z dr ,
\end{equation}
whereas the torque from the wall surface of the cylinder is
\begin{equation}
    \tau_{2} = \int_{-z_{end}}^{z_{end}} \frac{2\pi}{\mu_0} r_{end}^2 B_\theta B_r dz \; .
\end{equation}
Given the odd symmetry of $B_\theta$ about the $z=0$ middle plane shown in figure~\ref{fig: Btheta contour rr nu 0.23Va}, a net total electromagnetic torque can be evaluated as $\tau_{z, EM} = 2\tau_1 + \tau_2$.

\begin{figure*}[!htbp]
  \centering
  \includegraphics[width=0.8\textwidth]{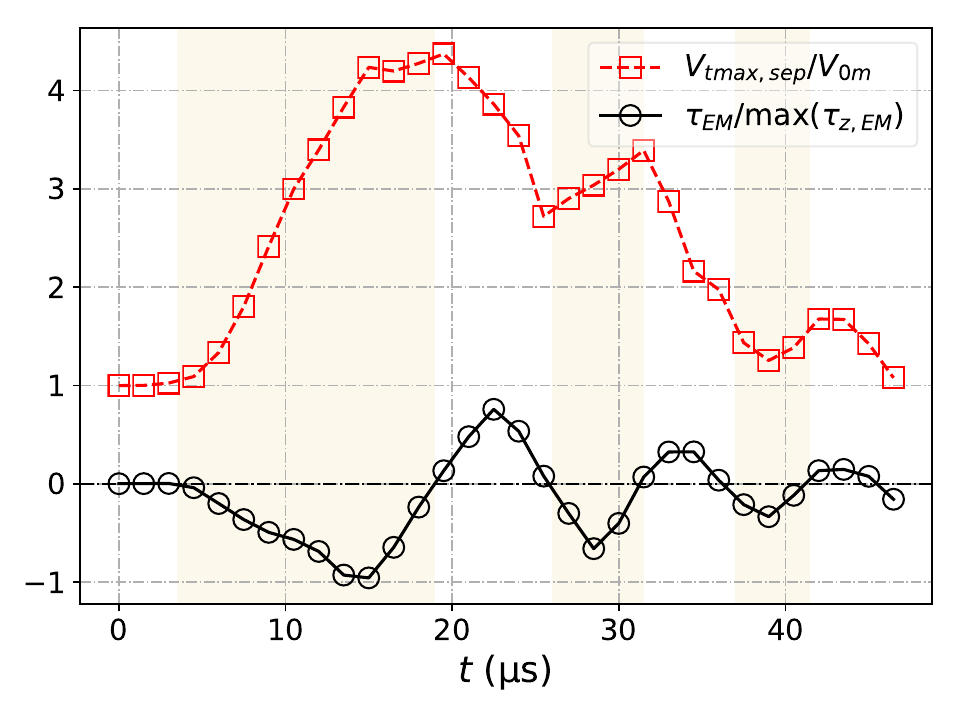}
  \caption{The time history of the normalized maximum toroidal velocity near the separatrix between $r=r_s$ and $r_{end}=1.3r_s$ in the $z=0$ middle plane (red) and the electromagnetic torque acting on the cylinder inside the volume specified by  $0 \leq r \leq r_{end}=1.3 r_s$ and $-z_{end} \leq z \leq z_{end}=1.6\mathrm{m}$ (black) for the initial rigid rotation case, with $\tau_{EM} < 0$ shaded yellow. The simulation is same as the case shown in figure~\ref{fig: RR results}.}
  \label{fig: torque RR}
\end{figure*}

The electromagnetic torque $\tau_{z,EM}$ acting on the cylinder enclosing the FRC region during compression is obtained for $r_{end}=1.3r_s$ and $z_{end}=1.6 \mathrm{m}$ (figure~\ref{fig: torque RR}).  The $B_z>0$ in the open field line region corresponds to an initial co-rotating flow $V_t>0$ that is counter to the $\theta$ direction in our simulations, and $\tau_{EM}=-\tau_{z,EM}<0$ highlighted in yellow in figure~\ref{fig: torque RR} thus represents a deceleration torque against the flow in the co-rotating direction. The electromagnetic torque $\tau_{EM}$ acts to decelerate the rotation as the  velocity increases and accelerate it as the velocity decreases.

The electromagnetic torque oscillation in figure~\ref{fig: torque RR} also indicates an exchange between kinetic and magnetic energies, which correlates with the twisting and unwinding of magnetic field lines during the compression process. The amplitude of the torque goes down as the compression becomes less dynamic. The energy exchanges primarily between the toroidal components $B_\theta$ and $u_\theta$, with minimal impact on the compression process in the poloidal plane. The torque modulates the variation of toroidal velocity and produces a pronounced flow shear near the separatrix, where substantial plasma outside the separatrix rotates in the opposite direction with $V_t<0$ (figure~\ref{fig: vt slice nu rr}). The rotation reversal continuously decelerates the velocity near the separatrix even as the torque $\tau_{EM}$ oscillates down to a negligible magnitude. In this way, the electromagnetic torque is able to divert away the  toroidal flow effects on compression scalings.

\subsection{Results with initially sheared rotation}

\begin{figure*}[!htbp]
  \centering
  \includegraphics[width=0.8\textwidth]{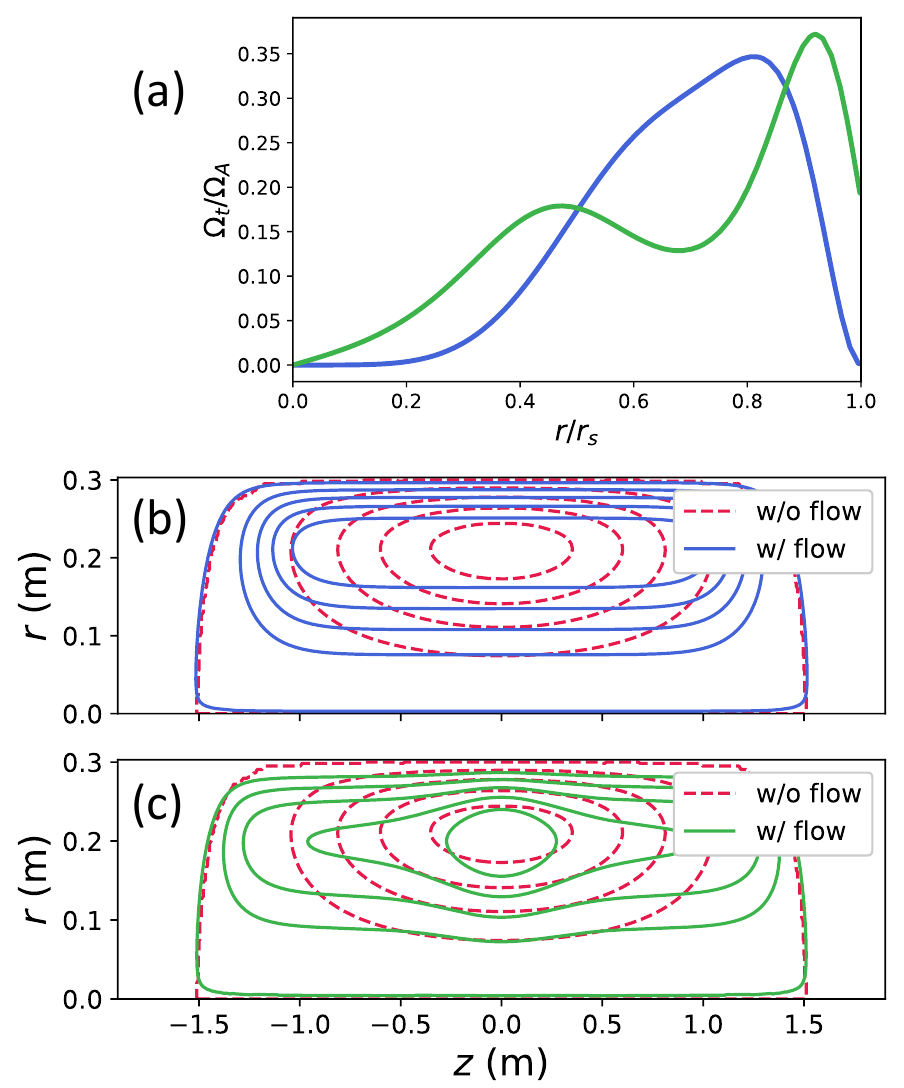}
  \caption{ (a) The single- (blue) and double-peaked (green) initial toroidal rotation profiles at the $z=0$ middle plane, and the corresponding equilibrium poloidal flux contours in presence of the (b) single- (blue) and the (c) double-peaked (green) rotation profiles in (a), in comparison to that of the equilibrium in absence flow (red).}
  \label{fig: eq flow}
\end{figure*}

\begin{figure*}[!htbp]
  \centering
  \includegraphics[width=0.95\textwidth]{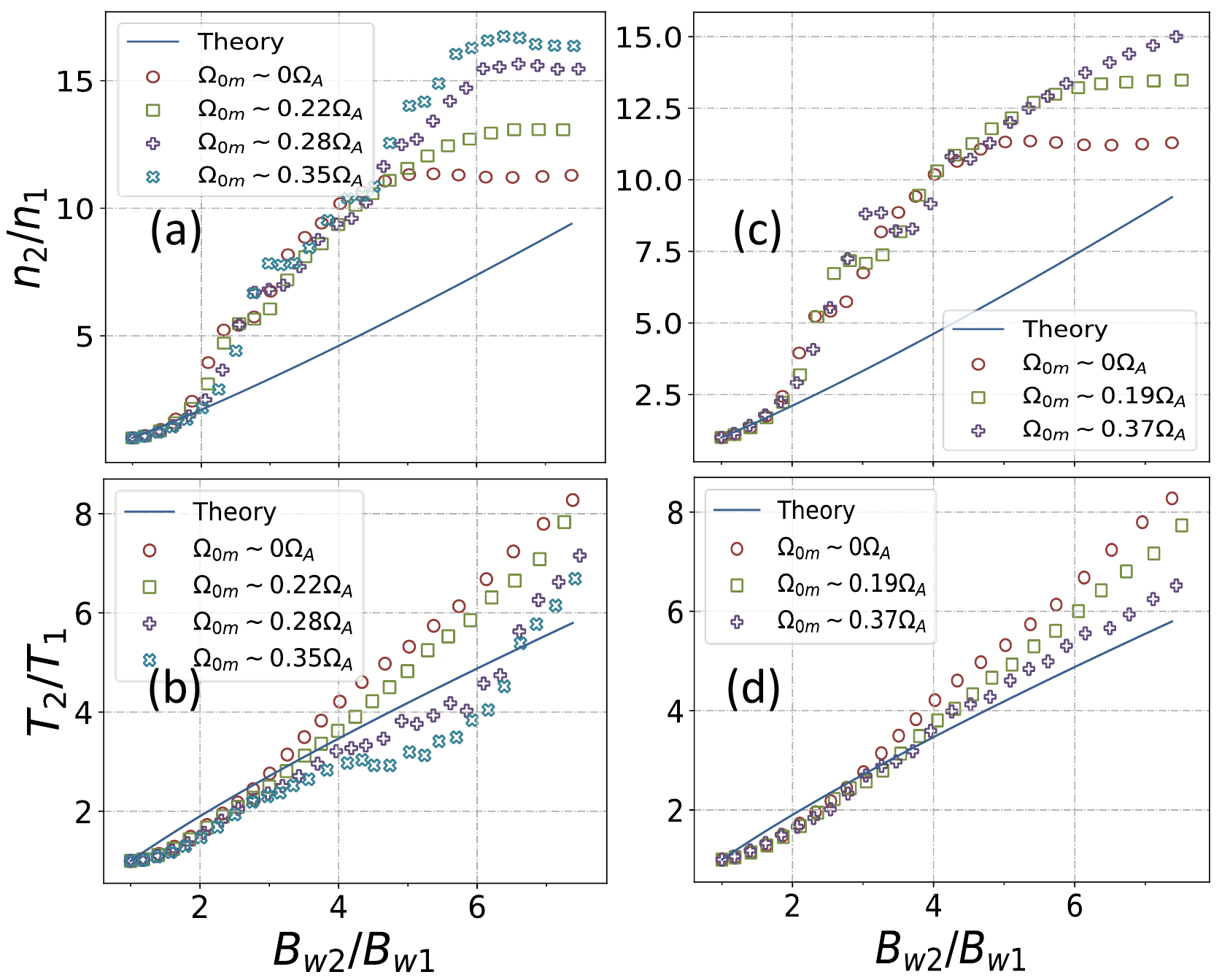}
  \caption{The compression ratio in (a) number density $n_2/n_1$ and (b) temperature $T_2/T_1$ in presence of initial single-peaked toroidal rotation profiles with various maximum rotation magnitude obtained from the Spencer theory (blue lines) and simulation results (symbols). The corresponding double-peaked rotation profile cases are shown in (c) and (d).}
  \label{fig: n t single dual nonuni}
\end{figure*}

\begin{figure*}[!htbp]
  \centering
  \includegraphics[width=0.8\textwidth]{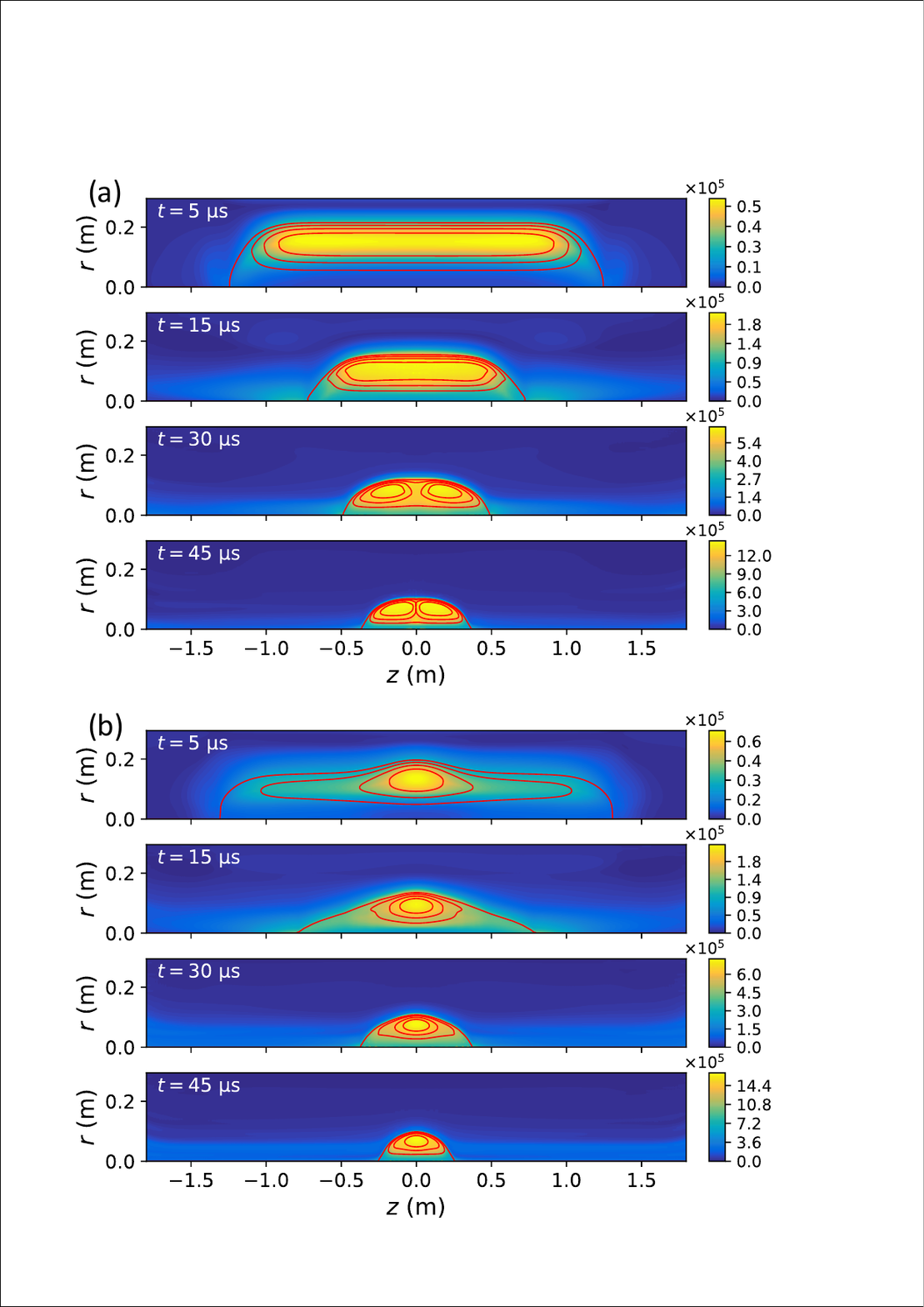}
  \caption{Pressure contours (colored) and poloidal flux contours (red) at different times during the compression for the same two cases shown in figure~\ref{fig: n t single dual nonuni} with (a) initially single-peaked toroidal rotation profile and $\Omega_{0m}\sim 0.35\Omega_A$, (b) initially double-peaked toroidal rotation profile and $\Omega_{0m}\sim 0.37 \Omega_A$.}
  \label{fig: p contour single dual nonuni}
\end{figure*}

Although the rigid rotation maybe straightforward and readily amenable to analytical solutions, significant flow shear is anticipated in more realistic FRC equilibria. Figure~\ref{fig: eq flow}(a) illustrates two additional types of initial toroidal rotation profiles, with their corresponding equilibrium poloidal flux displayed in figure~\ref{fig: eq flow}(b) and \ref{fig: eq flow}(c). Variations in rotation profiles can result in distinct distributions of equilibrium parameters. Specifically, in our simulations the single-peaked toroidal rotation profile elongates the internal poloidal flux contours, whereas the double-peaked toroidal rotation profile renders the core flux structure more oblate. These flow-profile induced variations in the initial equilibria can indirectly alter the magnetic compression scalings, as demonstrated in figure~\ref{fig: n t single dual nonuni}. The $\Omega_{0m}\sim 0.35 \Omega_A$ in the single-peaked rotation profile corresponds to $\alpha\sim 2$, and in the case of the double-peaked rotation profile the $\Omega_{0m} \sim 0.37 \Omega_A$ corresponds to $\alpha \sim 1.8$. The maximum density evolutions in both initial rotation profile cases exhibit similarities, which also agree with the case without initial flow until $B_{w2}/B_{w1} \gtrsim 5$. For the cases with the initial single-peaked rotation profile, however, a more significant temperature reduction is observed after $B_{w2}/B_{w1}>4$ and $\Omega_{0m} > 0.22\Omega_A$. This plasma temperature drop is reminiscent of the results in the FRX-C/LSM compression experiments~\cite{rej1992high}. The single-peaked rotation profile progressively elongates the core FRC equilibrium structure as the initial flow speed rises, leading to the formation of a thin and elongated current sheet. This elongation in the initial equilibrium structure is prone to field line tearing during compression, as shown in figure~\ref{fig: p contour single dual nonuni}. In contrast, the initial oblate core FRC structure associated with the double-peaked rotation profile inhibits the tearing mode and the subsequent plasma temperature drop during compression.

\begin{figure*}[!htbp]
  \centering
  \includegraphics[width=0.95\textwidth]{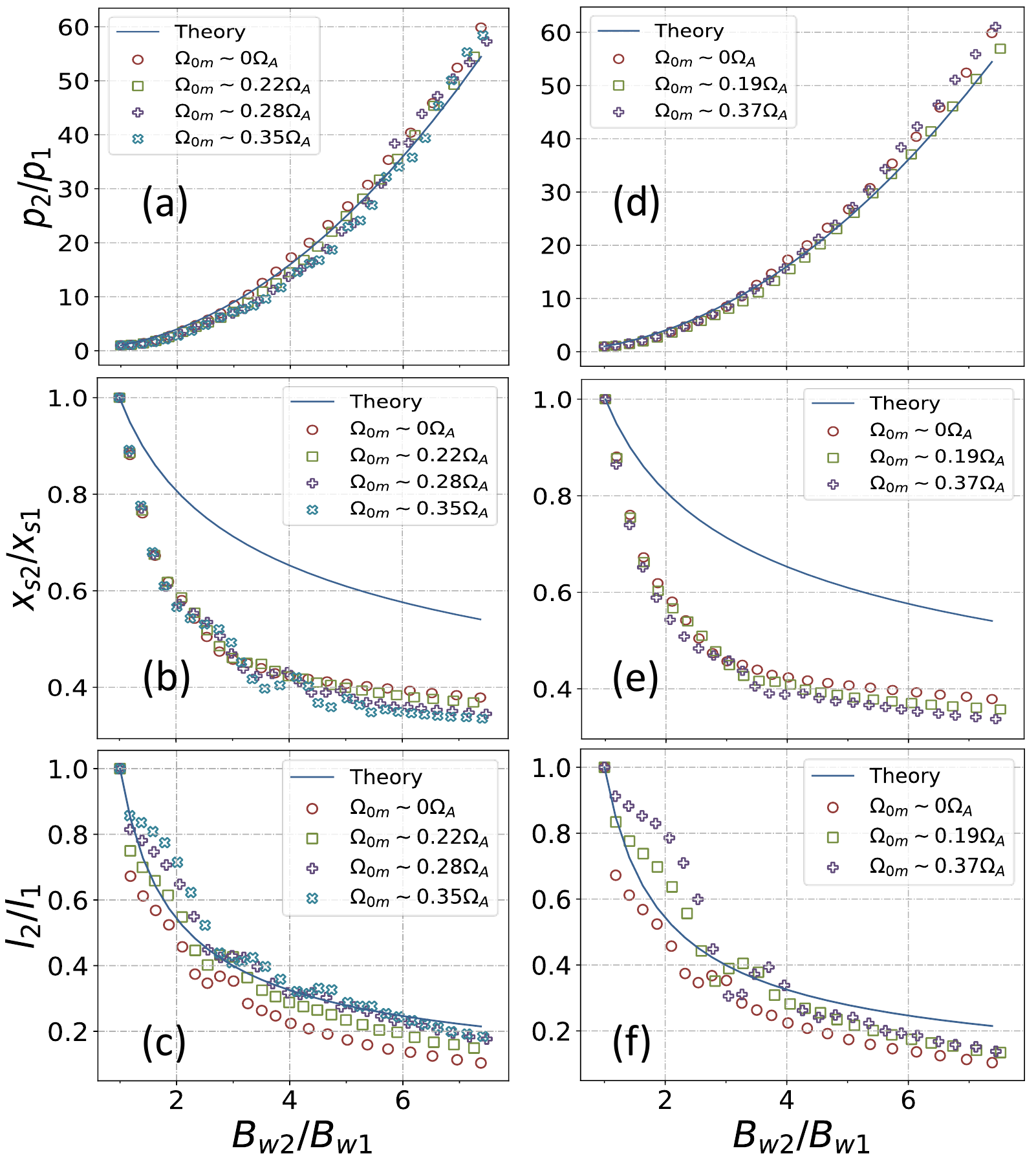}
  \caption{The compression ratio in (a) pressure $p_2/p_1$, (b) radius $x_{s2}/x_{s1}$ and (c) length $l_2/l_1$ in presence of initial single-peaked toroidal rotation profiles with various maximum rotation magnitude obtained from the Spencer theory (blue lines) and simulation results (symbols). The corresponding double-peaked rotation profile cases are shown in (d), (e) and (f). The simulations are same as the cases shown in figure~\ref{fig: n t single dual nonuni}.}
  \label{fig: p xs l single double nonuni}
\end{figure*}

\begin{figure*}[!htbp]
  \centering
  \includegraphics[width=0.9\textwidth]{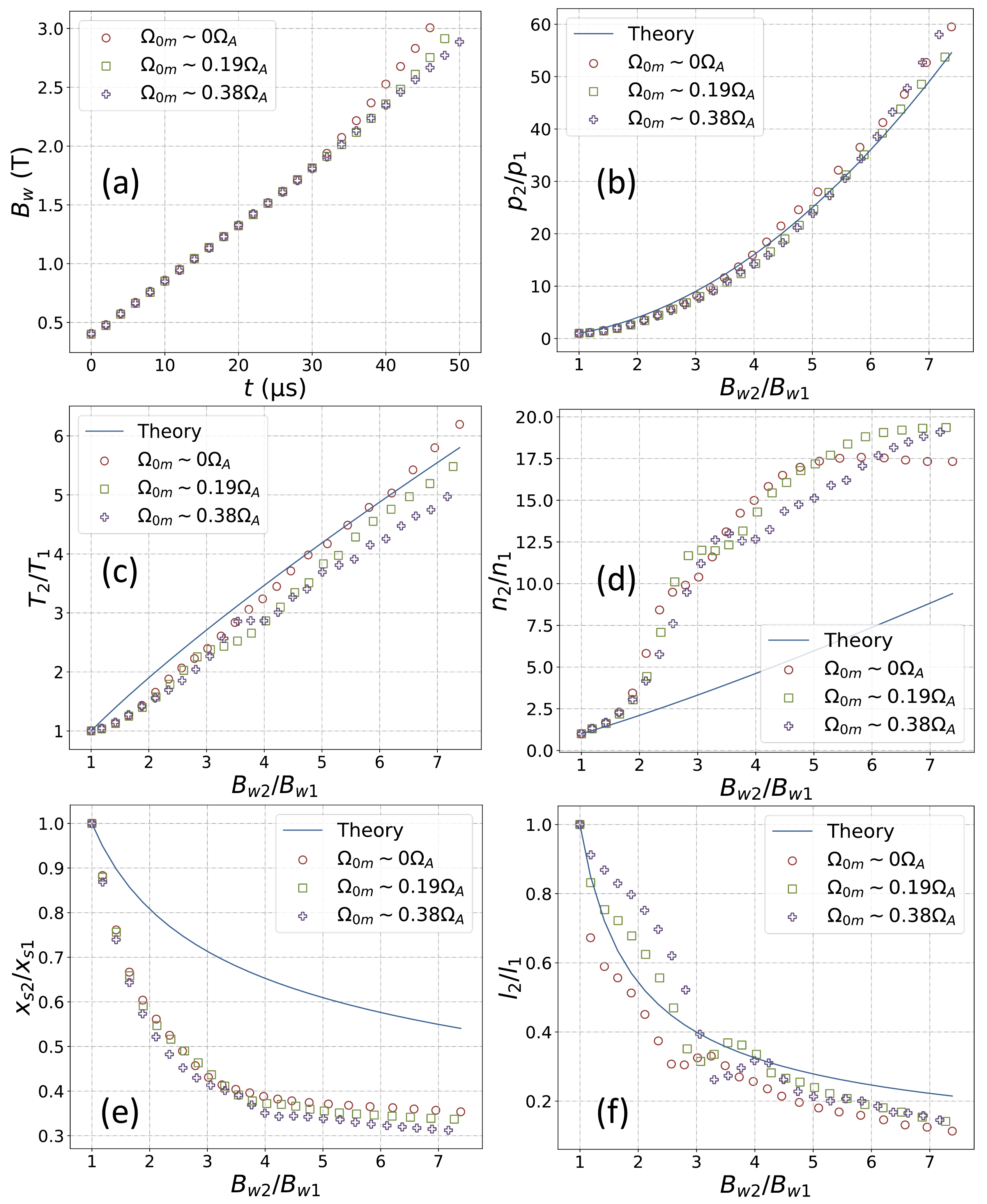}
  \caption{The comparisons between the Spencer's theory (blue lines) and simulation results with the initially double-peaked toroidal rotation profile. (a) The compression field $B_w$ variation with time, the (b) pressure ratio $p_2/p_1$, (c) temperature ratio $T_2/T_1$, (d) density ratio $n_2/n_1$, (e) radius ratio $x_{s2}/x_{s1}$ and (f) length ratio $l_2/l_1$ as functions of the magnetic compression ratio $B_{w2}/B_{w1}$ for different initial flow speeds. The simulation cases have an initially uniform density profile, differing from the cases in figures~\ref{fig: RR results}, \ref{fig: n t single dual nonuni} and \ref{fig: p xs l single double nonuni} where an initially peaked density profile is used.}
  \label{fig: spencer double uni}
\end{figure*}

The extent to which the pressure and radius scalings are affected by toroidal flow effects is not as pronounced as it is for the temperature scaling. Basically increasing the initial flow speed results in only slightly reduced FRC pressure and radius during compression. Such effects of the initial flows persist when considering the field line tearing during compression in cases with an initially single-peaked toroidal rotation profile, as illustrated in figure~\ref{fig: p xs l single double nonuni}. Contrary to the compressed FRC with the initial rigid rotation, the compressed FRCs exhibit slower axial contraction following the initial single-peaked and double-peaked rotation profiles, particularly in cases where tearing modes appear, as can be seen from figures~\ref{fig: p xs l single double nonuni} and \ref{fig: spencer double uni}. In scenarios featuring initial single-peaked and double-peaked rotation profiles, the flow effects on compression scalings under different initial density profiles show qualitative similarities except the density scaling. The density scaling is influenced by both the initial rotation and density profiles. For simulation cases with a uniform density profile (figure~\ref{fig: spencer double uni}), the density scaling remains generally lower compared to cases without initial flow before $B_{w2}/B_{w1} < 5$. By contrast in cases with initial peaked density profile where $n(r=0)/n_m \sim 0.5$ (figure~\ref{fig: n t single dual nonuni}), the density scaling aligns closely with the case without initial flow during $B_{w2}/B_{w1} < 5$.  Nevertheless, different density profiles do not significantly modify the overall flow effects on the compression scalings.

\begin{figure*}[!htbp]
  \centering
  \includegraphics[width=0.8\textwidth]{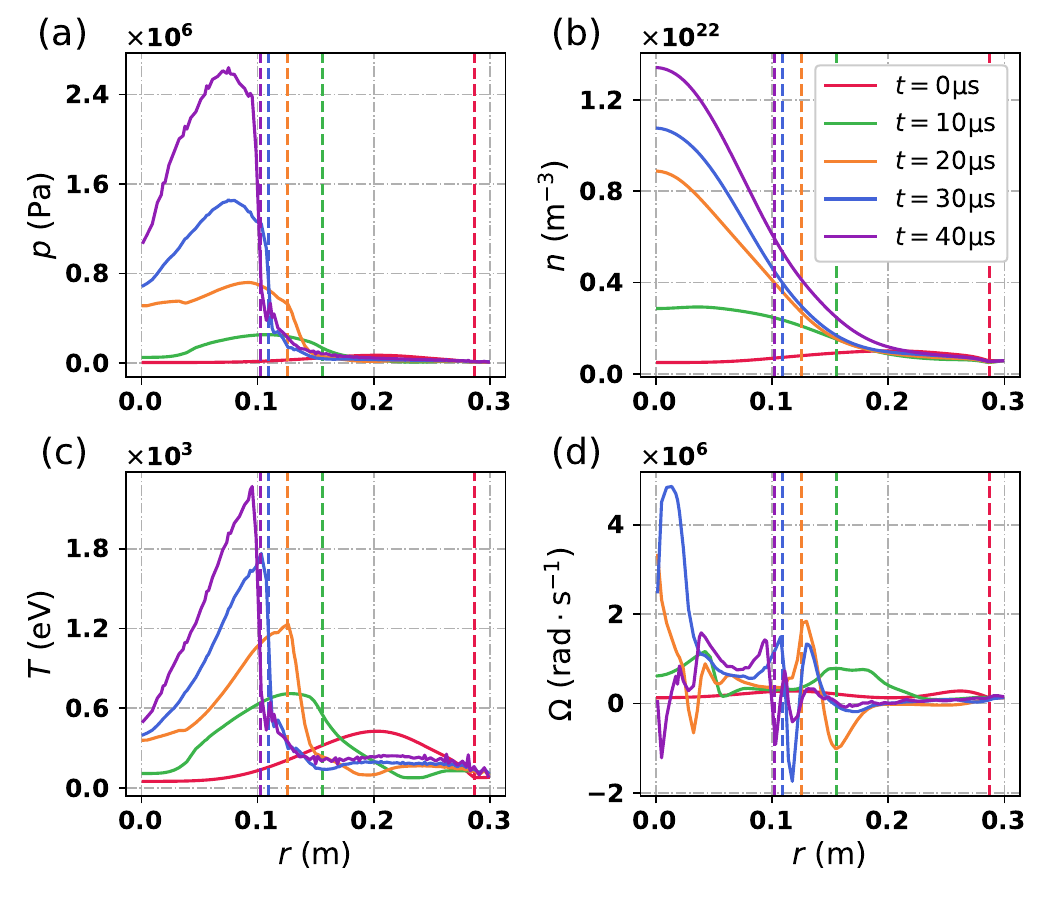}
  \caption{Radial profiles along the $z=0$ middle plane of the (a) pressure, (b) density, (c) plasma temperature and (d) toroidal rotation frequency at different times during compression respectively. The dashed lines indicate the locations of $r=r_s$ for the corresponding times. The simulation is same as the case shown in figure~\ref{fig: p contour single dual nonuni}(b).}
  \label{fig: nonuni dual 1d}
\end{figure*}

An enhanced toroidal rotation gradient near $r=r_s$ is induced during the compression, as can be seen from figure~\ref{fig: nonuni dual 1d} for the case with initial double-peaked rotation profile and $\Omega_{0m} \sim 0.37 \Omega_A$. The flow rotation frequency increases rapidly alongside the decreasing FRC radius. The rotation frequency rises to $\sim 6.4 \Omega_1$ at $t=20\mathrm{\mu s}$ and $r\sim0.13 \mathrm{m}$, and further to $\sim 15 \Omega_1$ at $t=30\mathrm{\mu s}$ and $r\sim0.02 \mathrm{m}$.  The magnitude of the toroidal rotation first increases and then decreases, which may partially explain the observed deviations in pressure and length scalings due to flow that is characterized by a rise and the subsequent fall during compression, as shown in figure~\ref{fig: spencer double uni}. Throughout the compression process, the pressure profile peaks around the $r_o$, and the temperature reaches its maximum between the $O$-point and $r_s$, whereas the maximum density accumulates at $r=0$ after $t>10\mathrm{\mu s}$. The discrepancy in the peak locations of the pressure, temperature and density profiles may account for their differences from the Spencer theory predictions. The steep gradient of the pressure and temperature profiles within $r_s$ can be attributed to the adiabatic limit of equation~(\ref{eqn: mhd temperature}) in the absence of thermal conduction and due to a slower compression rate than the maximum acoustic speed. Despite a wide range of variation in the toroidal rotation profile during compression, the shapes of the pressure, density and temperature profiles at the middle plane do not undergo fundamental changes compared to the case in absence of initial flow.

\section{Discussion and conclusions}
\label{sec: Discussion and conclusions}

Investigations into the toroidal flow effects on the magnetic compression scalings have been conducted. The MHD simulations employing the NIMROD code, which spans a range of initial rotation observed in the majority of FRC experiments and simulations, are compared with the simulation results and the Spencer scaling laws for static initial equilibriums. Various representative initial rotation and density profiles have also been considered. Basically the initial flow can affect the compression scalings through two mechanisms: one is the alteration of the compression process itself, resulting in a reduced compressional heating efficiency; the other is through reshaping the equilibrium  distributions, thereby indirectly influencing the scalings particularly on the temperature, density and length evolutions.

In the presence of initial rotation, the pressure scaling remains essentially same in comparison to the cases in absence of initial flow. The initial flow does tend to lower the temperature in the compressed state. A more significant temperature drop during the compression, reminiscent of the experimental results of the FRX-C/LSM device~\cite{rej1992high}, can be induced by the field line tearing which is more likely to take place when the flow effects elongate the core flux contours of the initial FRC equilibrium. Flow effects on the density scaling rely on both the initial flow and density profiles. Generally speaking, with initial flow the density scaling is slightly lower before $B_{w2}/B_{w1} < 5$, but becomes higher after $B_{w2}/B_{w1}>5$. The radius scaling can be influenced by flow effects with a deviation from the initially static case less than 15\%. The length scaling demonstrates an increase and it shows a faster axial contraction when $B_{w2}/B_{w1} < 3$ with the initially rigid rotation. In summary, for initially rotating FRCs, the pressure and the radius scalings are nearly the invariant  from the Spencer scalings for the magnetic compression ratio up to $B_{w2}/B_{w1}\sim 6$.

Such an invariance has been found a natural consequence of the dynamic angular momentum conservation due to the generation of a significant electromagnetic torque during the compression process, which acts on the plasma to oppose the acceleration of its toroidal rotation in general. A pronounced flow shear induced by the electromagnetic torque can continue to slow down the plasma rotation near the separatrix, even after the electromagnetic torque decays to negligible level. The electromagnetic torque oscillations during the compression indicate the twisting and unwinding of magnetic field lines, suggesting a dynamic exchange between kinetic and magnetic energies, which primarily involves the toroidal components $u_\theta$ and $B_\theta$, and has minimal influence on the magnetic compression dynamics within the poloidal plane. The generation of the $B_\theta$ by dynamo effects and the corresponding electromagnetic torque signifies the dynamic nature of compression missing in the 1D adiabatic model, as well as a key difference between the simulation and the quasi-static theoretical approaches.

The initial toroidal flow is expected to influence the FRC stability during compression. For example, toroidal flow is believed to excite the interchange modes driven by centrifugal force, particularly the $n=2$ rotational instability. On the other hand, our simulations with initial toroidal flow consistently show substantial $B_\theta$ and dynamically evolving rotation profiles with strong flow shear, which is generally considered beneficial for the FRC stability. Whereas the focus here is on the compression scalings, the stability of the compressed FRCs, particularly in presence of the evolving flow profiles, is an important topic that is planned for future investigation.

\section{Acknowledgement}

This work was supported by the National Key Research and Development Program of China (Grant No. 2019YFE03050004), the National Natural Science Foundation of China (Grant No. 51821005), and the U.S. Department of Energy (Grant No. DE-FG02-86ER53218). The computing work in this paper was supported by the Public Service Platform of High Performance Computing by Network and Computing Center of HUST. The authors are very grateful for the supports from the NIMROD team and the J-TEXT team.

\section*{References}
\normalsize
\bibliography{references}
\clearpage


\clearpage

\clearpage

\clearpage

\clearpage

\clearpage

\clearpage

\clearpage

\clearpage

\clearpage

\clearpage

\clearpage

\clearpage

\end{document}